\documentclass[12pt]{article}
\usepackage{epsfig}
\usepackage{amssymb,amsmath}

\newcommand{\bea}{\begin{eqnarray}}
\newcommand{\eea}{\end{eqnarray}}
\def\tr{\mathrm{tr}}
 \makeatletter
\def\alt{\mathrel{\mathpalette\gl@align<}}
\def\agt{\mathrel{\mathpalette\gl@align>}}
\def\gl@align#1#2{\lower.6ex\vbox{\baselineskip\z@skip\lineskip\z@
\ialign{$\m@th#1\hfil##\hfil$\crcr#2\crcr\sim\crcr}}} \makeatother

\begin{document}

\vspace*{1.0cm}

\begin{center}
\baselineskip 20pt {\Large\bf Type II Seesaw  and the PAMELA/ATIC
Signals}
\vspace{1cm}

{\large Ilia Gogoladze$^{a,}$\footnote{E-mail:
ilia@physics.udel.edu\\ \hspace*{0.5cm} On  leave of absence from:
Andronikashvili Institute of Physics, GAS, 380077 Tbilisi, Georgia.}, Nobuchika Okada$^{b,}$\footnote{ E-mail: okadan@post.kek.jp} and
Qaisar Shafi$^{a,}$ } \vspace{.5cm}

{\baselineskip 20pt \it
$^a$Bartol Research Institute, Department of Physics and Astronomy, \\
 University of Delaware, Newark, DE 19716, USA \\
\vspace{2mm}
$^b$
Theory Group, KEK, Tsukuba 305-0801, Japan }
\vspace{.5cm}

\vspace{1.5cm} {\bf Abstract}
\end{center}

\noindent We discuss how the cosmic ray signals reported by the PAMELA and ATIC/PPB-BETS experiments may be understood
in a Standard Model (SM) framework supplemented by type II seesaw and
 a stable SM singlet  scalar boson as dark matter. A particle physics explanation of the
'boost' factor can be provided by including an additional SM singlet scalar field.
\thispagestyle{empty}

\newpage

\addtocounter{page}{-1}


The PAMELA experiment has reported a significant positron excess
over the expected background without a corresponding increase in the
flux of anti-protons \cite{Adriani:2008zr}.
Their measurement seems to be consistent within the error bars
with results previously reported by HEAT \cite{Barwick:1997ig}
and AMS \cite{Aguilar:2007yf}.
More recently, the ATIC experiment \cite{chang:2008zz}
(see also PPB-BETS \cite{Torii:2008xu}) has reported an appreciable
flux of electrons and positrons at energies of around 100 - 800 GeV,
which also appears to be significantly higher than the expected
background at these energies.
While  pulsars and/or other nearby astrophysical
sources may account for the PAMELA results alone \cite{Aharonian(1995)},
 a unified understanding of both the PAMELA and ATIC/PPB-BETS
measurements based on such  sources appears to be  more challenging.

A particle physics inspired explanation of both the PAMELA and
ATIC/PPB-BETS results in terms of dark matter physics necessitates
a suitable extension of the  SM framework.
For instance, the dark matter could be a stable elementary particle
with suitably chosen mass which primarily self annihilates into leptons
(leptophilic) through new interactions \cite{dm}.
Depending on the details, this scenario  may, in addition,  invoke
a `boost' factor which could  either have  an astrophysical origin  (large inhomogeneities in the dark matter distribution),
or have a particle physics origin such
as Sommerfeld enhancement \cite{sommerfeld}.
Alternatively, one could assume that the dark matter is not entirely stable
but extremely long-lived, with a lifetime  $\sim 10^{26}$ sec \cite{Yin:2008bs}.

In this letter we  offer what we believe is a very
simple extension of the SM according to which  stable dark matter annihilates  primarily into leptons.
The dark matter particle in our scheme is
 a SM singlet boson $D$ \cite{darkon, elastic},
and its stability comes  from an unbroken $Z_2$ symmetry under
which it carries negative parity.
The leptophilic nature of $D$ arises from its interactions
with the SU(2) triplet scalar fields which are introduced
to accommodate the observed neutrino oscillations \cite{NuData}
via the type II seesaw mechanism \cite{seesawII}.
In the minimal version the model requires a 'boost' factor, of
order $10^3$ or so, which should have an astrophysical origin \cite{Hooper:2008kv}.
We then proceed to show how the presence of an additional SM singlet
scalar field can provide a particle physics origin for
this 'boost' factor based on the Breit-Wigner enhancement of
dark matter annihilation \cite{Ibe:2008ye}.
%

The particle content relevant for our discussion in this paper
 is summarized  in Table~1.
The SM singlet scalar is assigned  an odd  $Z_2$ parity which makes it stable and a suitable  dark matter candidate.
It is often useful to explicitly express the triplet scalar
 by three complex scalars (electric charge neutral,
 singly charged and doubly charged scalars):
\bea
 \Delta=\frac{\sigma^i}{\sqrt{2}}
   \Delta_i=\left(
 \begin{array}{cc}
    \Delta^+/\sqrt{2} & \Delta^{++}\\
    \Delta^0 & -\Delta^+/\sqrt{2}\\
 \end{array}\right) .
\eea

\begin{table}[t]
\begin{center}
\begin{tabular}{c|cc|c}
           & SU(2)$_L$ & U(1)$_Y$ & $Z_2$  \\
\hline
$ \Delta $ & {\bf 3}  & $+1$      & $+$  \\
$ \ell_L^i   $ & {\bf 2}  & $-1/2$    & $+$  \\
$ H      $ & {\bf 2}  & $+1/2$    & $+$  \\
\hline
$ D      $ & {\bf 1}  & $ 0  $    & $-$  \\
\end{tabular}
\end{center}
\caption{
Particle content relevant  for our discussion in this paper.
In addition to the SM lepton doublets $\ell_L^i$ ($i$ is the generation index)
 and the Higgs doublets $H$,
 a complex scalar $\Delta$ and a real scalar $D$ are introduced.
The triplet scaler $\Delta$ plays the key role in type II seesaw mechanism,
 while $D$ is the dark mater candidate.
}
\end{table}

The scalar potential relevant for the type II seesaw is given by
\bea
 V(H, \Delta) &=&
 -m_H^2 (H^\dagger H)
  + \frac{\lambda}{2} (H^\dagger H)^2 \nonumber\\
&+& M_\Delta^2 \tr \left( \Delta^\dagger \Delta \right)
 + \frac{\lambda_1}{2} \left(\tr \Delta^\dagger \Delta \right)^2
 + \frac{\lambda_2}{2}\left[
  \tr \left( \Delta^\dagger \Delta \right)^2
 -\tr \left(\Delta^\dagger \Delta \Delta^\dagger \Delta \right)
 \right] \nonumber\\
&+&
 \lambda_4 H^\dagger H \; \tr \left(\Delta^\dagger\Delta\right)
 + \lambda_5 H^\dagger
 \left[\Delta^\dagger, \Delta\right] H
 + \left[ 2 \lambda_6 M_\Delta
    H^T i\sigma_2 \Delta^\dagger H +{\rm h.c.} \right],
 \label{H-Delta-Potential}
\eea
where  the coupling constants $\lambda_i$
 are taken to be real without loss of generality.
The triplet scalar has a Yukawa coupling with the lepton doublets given by
\bea
{\cal L}_\Delta &=&
 -\frac{1}{2}\left(Y_\Delta\right)_{ij}
  \ell_L^{Ti}\, \mathrm{C} \, i\, \sigma_2\,  \Delta \, \ell_L^j +{\rm h.c.}\nonumber \\
  &=&
 -\frac{1}{2} \left(Y_\Delta\right)_{ij} \,
  \nu^{Ti}_L \, \mathrm{C} \, \Delta^0 \, \nu_L^j
  +  \frac{1}{\sqrt{2}} \, \left(Y_\Delta\right)_{ij} \,
  \nu^{Ti}_L \, \mathrm{C} \, \Delta^+ \ e_L^j
  + \frac{1}{2} \left(Y_\Delta\right)_{ij} \, e^{Ti}_L \,
  \mathrm{C} \, \Delta^{++} \  e_L^j  +  \text{h.c.}  ,
\label{Yukawa}
\eea
where $\mathrm{C}$ is the Dirac charge conjugate matrix
 and $\left(Y_\Delta\right)_{ij}$ denotes elements
 of the Yukawa matrix.

A non-zero vacuum expectation value (VEV) of the
 Higgs doublet induces a tadpole term for $\Delta$
 through the last term in Eq.~(\ref{H-Delta-Potential}).
A non-zero VEV of the triplet Higgs is thereby generated,
 $\langle \Delta_0 \rangle = v_\Delta/\sqrt{2} \sim \lambda_6
 v^2/M_\Delta$ ($v=246$), which leads to the neutrino mass
 from Eq.~(\ref{Yukawa}):
\bea
  M_\nu = \left(Y_\Delta\right)_{ij} \langle \Delta_0 \rangle .
\eea

Note that the triplet Higgs VEV contributes to the weak boson
 masses and alters the $\rho$-parameter from the SM prediction,
 $\rho \approx 1$, at tree level.
The current precision measurement \cite{PDG}
 constrains this deviation to be in the range,
 $ \Delta \rho =\rho -1 \simeq
 \langle \Delta \rangle/v  \lesssim 0.01$,
 so that $\lambda_6 \lesssim 0.01 M_\Delta/v$.
This constraint is especially relevant if we take  $M_\Delta = {\cal
O}$(100 GeV), in which case the region $\lambda_6 \gtrsim 0.01$ is excluded.

The scalar potential relevant for  dark matter physics is given by
\bea
V(H, \Delta, D) &=&
 \frac{1}{2} m_0^2 D^2 + \lambda_D D^4
 + \lambda_H D^2  (H^\dagger H)
 + \lambda_\Delta D^2 \tr (\Delta^\dagger \Delta)  \nonumber \\
&=& \frac{1}{2} m_D^2 D^2 + \lambda_D D^4
 + \lambda_H v D^2 h  + \frac{\lambda_H}{2} D^2 h^2 \nonumber \\
&+& \lambda_\Delta D^2
\left(
 \sqrt{2} v_\Delta \Re[\Delta_0] + |\Delta_0|^2 + |\Delta^+|^2 + |\Delta^{++}|^2
 \right),
\label{H-Delta-D-Potential}
\eea
where $m_D^2 = m_0^2 + \lambda_H v^2 + \lambda_\Delta v_\Delta^2$,
 and in the last equality the potential is expressed
 in terms of physical Higgs bosons ($h$).

We first investigate the relic abundance of the singlet dark matter,
 which is obtained by solving the following Boltzmann equation \cite{Kolb:1990vq},
\begin{eqnarray}
 \frac{d Y}{d x}
 =
 -\frac{\langle \sigma {\rm v} \rangle}{H x}s
 \left( Y^2 - Y_{\rm eq}^2 \right)~,
 \label{eq:Boltzmann}
\end{eqnarray}
 where $Y = n/s$ is the  ratio of
 the dark matter number density ($n$) to the entropy density of the universe
 ($s = 0.439 g_* m_D^3/x^3$), $g_* \sim 100$, and
 $x \equiv m_D/T$ ($T$ is the temperature of the universe).
The Hubble parameter is given by
 $H = 1.66 g_*^{1/2} m_D^2 m_{\rm PL}/x^2$,
 where $m_{\rm PL} = 1.22\times 10^{19}$ GeV is the Planck mass,
 and the dark matter yield in  equilibrium  is  $Y_{\rm eq} = (0.434/{g_*}) x^{3/2} e^{-x}$.
Solving the Boltzmann equation with the thermal averaged
 annihilation cross section $\langle\sigma {\rm v}\rangle$,
 we obtain the relic abundance of dark matter ($Y_\infty$).
To  a good accuracy, the solution of Eq.~(\ref{eq:Boltzmann})
 is approximately given by \cite{Kolb:1990vq}
\begin{eqnarray}
 \Omega h^2
 &=&
 \frac{1.07\times 10^9 x_f{\rm GeV}^{-1}}
 {\sqrt{g_*}m_{\rm PL}\langle\sigma {\rm v}\rangle} ,
\label{Omega}
\end{eqnarray}
 where $x_f = m_D/T_f$,  the freeze-out temperature
 for  dark matter, is given by  $x_f = \ln(X) - 0.5\ln(\ln(X))$,
 with $X = 0.038 (1/g_*^{1/2})m_{\rm PL} m_D \langle
 \sigma {\rm v}\rangle$.
If  the dark matter annihilation occurs in the s-wave
 at the non-relativistic limit, the thermal averaged
 annihilation cross section $\langle\sigma {\rm v}\rangle$
 is simply replaced by non-averaged one,
 $\langle\sigma {\rm v}\rangle = \sigma {\rm v}$.

In the following  we consider the case $m_D > M_\Delta,\  m_h$,
 where $m_h$ is the SM Higgs boson mass.
In this case, we find that the dominant dark matter annihilation process
 is $DD \to hh, \Delta^\dagger \Delta$
 through the quartic coupling $\lambda_H$ and $\lambda_\Delta$
 in Eq.~(\ref{H-Delta-D-Potential}). In  the non-relativistic limit
the cross section  is given by
\bea
 \sigma v = \frac{1}{16 \pi m_D^2}
 \left( \lambda_H^2 + 6 \lambda_\Delta^2\right).
\eea
For a given $m_D$, the annihilation cross sections is determined
 so as to satisfy the observed relic density of dark matter \cite{WMAP},
\bea
  \Omega_{DM} h^2 \simeq  0.1131.
\eea
For example, the following parameter set can reproduce
 the observed dark matter relic density:
\bea
 && m_D =1.3 \; {\rm TeV},  \nonumber \\
 && \lambda_H^2 + 6 \lambda_\Delta^2 =  0.16 ,
 \label{parameter-set}
\eea
 which leads to   $\langle \sigma v \rangle = 1.85 \times 10^{-9}$ GeV$^{-2}= 0.72$ pb.

A variety of experiments are underway  to directly detect
 dark matter particles  through  elastic scattering off nuclei.
The most stringent limit on the (spin-independent) elastic
 scattering cross section has been obtained
 by the recent XENON10 \cite{XENON} and CDMS II \cite{CDMS} experiments:
 $\sigma_{el}({\rm cm}^2) \lesssim 7 \times 10^{-44} - 5 \times 10^{-43}$,
 for a dark matter mass of 100 GeV$\lesssim m_{DM} \lesssim$ 1 TeV.
Since the singlet  $D$  can scatter off a nucleon
 through  processes mediated by the SM Higgs boson in the t-channel,
 a parameter region of our model is constrained by this current
 experimental bound.
The elastic scattering cross section for this process
 is estimated to be \cite{elastic}
\bea
  \sigma_{el} \sim 1.4 \times 10^{-45} ({\rm cm}^2)
  \times
  \left(\frac{\lambda_H^2}{0.1} \right)
  \left( \frac{1.3{\rm TeV}}{m_D} \right)^2
  \left( \frac{120{\rm GeV}}{m_h} \right)^4.
\eea
For the parameter set in Eq.~(\ref{parameter-set}),
 this cross section is two orders of magnitude smaller
 than the current bound, but could be within the reach
 of future experiments if $\lambda_H^2={\cal O}(0.1)$.

The dark matter in the halo of  our galaxy can  annihilate
 and produce high energy SM particles.
In the case of $D$ we obtain the triplet ($\Delta$) and
 the SM Higgs bosons through the same processes
 as in the early universe with the same annihilation cross section.
Thus, pairs of the  Higgs triplet  and the SM Higgs bosons
 are produced which  eventually decay into the lighter SM particles, and thus
 provide additional contributions to cosmic ray fluxes.
In this paper we assume $\lambda_H < \lambda_\Delta$ so that  the dark matter pair dominantly annihilates
 into the Higgs triplet of  type II seesaw.
There are two types of decay modes of the triplet Higgs boson.
One is into lepton pairs through the Yukawa coupling $Y_\Delta$
 which has a direct relation to the neutrino oscillation data
 through the type II seesaw mechanism.
The second decay mode contains  gauge bosons and SM Higgs boson pairs  and proceeds through
 the gauge interactions and the couplings $\lambda_{4,5,6}$.
Note that the decay amplitudes in the latter case are proportional
 to the small VEV of the triplet scalars, and hence the  Higgs triplet dominantly
 decay into lepton pairs
 unless the Yukawa coupling $Y_\Delta$ is very small
 ($Y_\Delta \lesssim v_\Delta/M_\Delta$ as a rough estimate).
Therefore, our model predicts that the cosmic rays originating
 from  dark matter annihilation in the halo are primarily leptons.
This is a remarkable feature when we consider the experimentally observed  cosmic ray positron/electron excess,
with  no corresponding  excess in the  cosmic ray anti-proton flux.

It has been argued  \cite{Hisano:2009rc} that the excess in  cosmic ray
 electron/positron fluxes observed by PAMELA and ATIC/PPB-BETS can be simultaneously explained
  through lepton pairs produced
 by  dark matter annihilation in the halo with suitable
 energy for the  primary leptons;
an $e^+ e^-$ pair each with  650 GeV of energy  produced
 through  pair annihilation with a  cross section of about 100 pb,
 or a  $\mu^+ \mu^-$ pair or a $\tau^+ \tau^-$ pair
 with about 1 TeV energy each produced by pair annihilation
 with a cross section of about 1000 pb.
Note that in order to explain the excess of cosmic rays,
 the dark matter annihilation cross section should be
 two or three orders of magnitude larger than the typical
 cross section $\sim$ 1 pb which yields  the correct relic abundance.
We simply assume that the difference  is provided by the so-called
`boost'  factor originating from the inhomogeneity of the dark matter
 distribution in the halo. A particle physics explanation for this `boost' factor will shortly be discussed.

\begin{figure}[]\begin{center}
\includegraphics[scale=0.6]{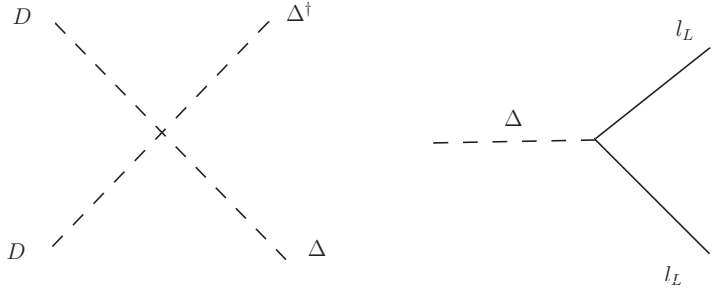}
\caption{Dark matter annihilation into pair of $\Delta$'s; $\Delta$ decay into pair of leptons.
}
\end{center}
\end{figure}

In our model a pair of $D$'s annihilates into a pair
of triplet higgs bosons which, in turn, produce a total
of four leptons (see Figure 1). In order to explain the PAMELA and ATIC/PPB-BETS signals, the dark matter mass should be roughly twice the observed positron energy of 650 GeV \footnote[1]{
The energy distribution of the final state leptons is not monochromatic
 since they are produced by the decay of the boosted Higgs triplet bosons.
Thus, more precisely,  the dark matter mass required to fit
 the PAMELA and ATIC/PPB-BETS data would be slightly larger.
The effect of this energy distributions of the final state leptons
 is reflected in the cosmic ray electron flux which can  be a key
 to sort out dark matter models accounting for the excess reported by
 PAMELA and ATIC/PPB-BETS  \cite{Chen}. }.

If the decay product is mainly  $e^+ e^-$,  the required dark matter mass is
 around 1.3 TeV, while $m_D \sim 2$ TeV is needed if  $\mu^+ \mu^-$ pair
 or $\tau^+ \tau^-$ pair dominates the annihilation channel.
There is an interesting implication of this because of
 the underlying  type II seesaw mechanism.
The primary leptons are produced by the triplet Higgs boson decay,
 so that the final state lepton flavor has a direct relation
 with the Yukawa coupling $Y_\Delta$ and hence also with  the neutrino mass matrix.
The normal hierarchical mass spectrum of neutrinos predicts
 that  $\mu^+ \mu^-$ and $\tau^+ \tau^-$ pairs are the dominant
 decay channels, while  comparable amounts
 of $e^+ e^-$,  $\mu^+ \mu^-$ and $\tau^+ \tau^-$ pairs are produced
 in the inverted-hierarchical neutrino mass spectrum.
A precise measurement of the energy dependence of the positron/electron flux
 may allow us to distinguish these two neutrino mass spectra
 because the flux of primary $e^+ e^-$ pair shows a sharp drop
 at the maximum cosmic ray energy (half of the dark matter mass).

As previously stated the dark matter annihilation cross section required  to account for
 the PAMELA and ATIC/PPB-BETS data should be a few orders  of magnitude
 larger than the one suitable for obtaining the correct relic abundance.
In the above discussion, we simply assumed that the boost factor
 of  astrophysical origin provides the required degree of  enhancement of the cross section.
It would be more interesting if the boost factor emerges
 as a result of some mechanism from particle physics.
In the following, we show that a simple extension of our model
 can indeed provide such a boost factor.

We consider the Breit-Wigner enhancement of dark matter annihilation
 proposed in \cite{Ibe:2008ye}.
In this mechanism, the dark matter pair annihilation
 in the present universe occurs trough  an  s-channel process
 mediated by a state with mass very close to but slightly
 smaller than  twice  the dark matter mass.
Although the same process is also relevant for  dark matter
 annihilation in the early universe, a relative velocity
 between annihilating dark matters at the freeze-out time
 is not negligible, and the total anergy of two annihilating $D$'s
 is pushed away from the s-channel resonance pole.
As a result, we can obtain a relatively large suppression of the annihilation
 cross section at the freeze-out time compared to the one at present.

To implement this scenario we introduce a $Z_2$-parity even
 real scalar $(S)$ which is a singlet under  the SM gauge group.
We focus on the following scalar potential:
\bea
 V(S, D, \Delta) = \frac{1}{2} M_S^2 S^2
           + \lambda_1 M_S S D^2
           + \lambda_2 M_S S \tr(\Delta^\dagger \Delta).
\eea
We assume $M_S > M_\Delta$ and also that other  couplings
 involving $S$ are negligibly small.

\begin{figure}[]\begin{center}
\includegraphics[scale=0.6]{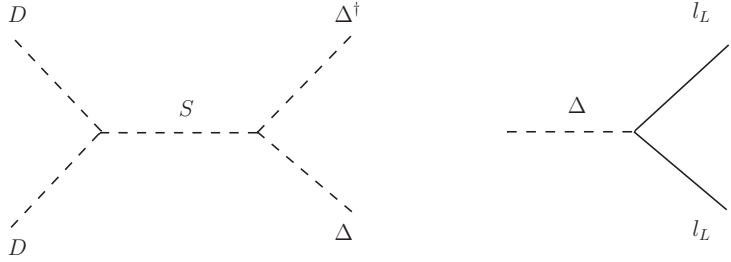}
\caption{Dark matter annihilation into $\Delta$'s
mediated by S; $\Delta$ decay into pair of leptons.}
\end{center}
\end{figure}

Next  consider the annihilation process mediated
 by the singlet, $DD \to S \to \Delta^\dagger \Delta$ (see Figure~2).
The annihilation cross section times relative velocity
 in the zero-velocity limit is calculated to be
\bea
  \sigma v|_{v \to 0} = \frac{8 \lambda_1^2 M_S^2}{(4 m_D^2 - M_S^2)^2 + M_S^2
  \Gamma_S^2} \frac{\tilde{\Gamma}_S}{2 m_D},
\eea
where the total decay width of the $S$ boson is given by
 $\Gamma_S = (3 \lambda_2^2/16 \pi) M_S$,
 and $\tilde{\Gamma}_S= \Gamma_S(M_S \to 2 m_D)$.
According to Ref.~\cite{Ibe:2008ye}, we introduce
 two small parameters ($0 < \delta \ll 1$ and $\gamma \ll 1$):
\bea
  M_S^2 = 4 m_D^2 (1-\delta), \; \;
  \gamma = \frac{\Gamma_S}{M_S}= \frac{3 \lambda_2^2}{16 \pi},
\eea
so that  the cross section formula is rewritten as
\bea
  \sigma v|_{v \to 0} \simeq \frac{2 \lambda_1^2}{m_D^2}
    \frac{\gamma}{\delta^2 + \gamma^2}.
\eea

For $\delta, \gamma \ll 1$, the parameters are set to be
 very close to the $S$-resonance pole, and  the relative
 velocity of annihilating dark matter in the early universe,
 $v \sim {\cal O}(0.1)$, causes a large suppression
 of the annihilation cross section  ($v \gg \delta, \gamma$).
In \cite{Ibe:2008ye}, this suppression factor
 (in other words, the inverse of the boost factor $BF^{-1}$)
 is estimated as $BF^{-1} \sim 10 \times {\rm Max}[\delta, \gamma]$.
In order to account for the excess in the PAMELA and ATIC/PPB-BETS
 experiments we impose $ \sigma v|_{v \to 0} \sim 1000$ pb,
 while $BF^{-1} \sim 1000$ to obtain the correct relic abundance
 of the dark matter, $ \langle \sigma v \rangle \sim 1$ pb.
It is in fact easy to satisfy these conditions
 by tuning the model parameters.
For example, if we take $\lambda_1 \sim 0.01$ and $\lambda_2 \sim 0.04$,
 these conditions are satisfied with $\delta \sim \gamma \sim 10^{-4}$.
In this case, it is not necessary for the process, $DD \to hh, \Delta^\dagger \Delta$,
 examined before, to be the dominant annihilation
 process, so that we take $\lambda_H^2 + 6 \lambda_\Delta^2 <  0.16$
 (see Eq.~(\ref{parameter-set})).

In summary, we have proposed a simple extension of the SM
 to accommodate both non-zero  neutrino masses and the observed dark matter
 in the  universe.
An SU(2)$_L$ triplet scalar with  unit hypercharge and
 a $Z_2$-parity odd real scalar singlet
 are introduced.
The triplet scalar implements type II seesaw
 while the singlet scalar $D$ is the dark matter candidate.
The relic density of $D$ depends on
  the annihilation process $DD \to \Delta^\dagger \Delta$,
 and  we have identified  the desired  parameter region.
The singlet dark matter particles in the halo of our galaxy
  annihilate into the triplet scalars  whose subsequent decay produces  lepton pairs.
Assuming a suitable astrophysical boost factor, these
 leptons can account for the excess in cosmic-ray
 positron/electron fluxes with a dark matter mass of  around 1 TeV.
Because of the nature of type II seesaw, the triplet Higgs bosons
 have no direct coupling with quarks, so that there is no sizable
 contribution to the cosmic-ray anti-proton flux.
We have also proposed a further extension of the model
 by introducing a $Z_2$-parity even real SM scalar singlet $S$.
In this case, the dark matter annihilation into the Higgs triplet bosons
 proceeds  through an s-channel process mediated by the singlet $S$.
With appropriate tuning of  parameters,
 the annihilation cross section of  dark matter in the present
 universe is enhanced through the Breit-Wigner enhancement mechanism
 \cite{Ibe:2008ye},
 while keeping the annihilation cross section in the early universe
 to be of  the right size ($\sim 1$ pb).
This extension can account for the  the cosmic ray
 positron/electron excess  without invoking an astrophysical boost factor.

Finally, we offer some concluding remarks. First, our model has
important implications
 for the SM Higgs boson mass.
As shown in \cite{Gogoladze:2008gf},
 the SM Higgs boson mass bounds obtained from imposing vacuum stability
 and perturbativity of the quartic Higgs coupling
 can be dramatically altered in the presence of type II seesaw.
 In particular, the Higgs boson mass window with type II seesaw can encompass
 mass regions otherwise not allowed. Indeed, the Higgs boson mass can
 even coincide with  the current experimental lower bound
 of  $m_H =114.4$ GeV \cite{Schael:2006cr}.
Second, the seesaw Higgs triplet is lighter than the mass ($\sim$ TeV) of
 the singlet dark matter particle.
A Higgs triplet boson this light should be produced in hadron colliders,
 especially the Large Hadron Collider \cite{LHCtypeII}.
In particular, the doubly-charged scalar may provide
 a clean signature through its decay into a pair of
 same sign charged leptons.

\section*{Acknowledgments}
We would like to thank Shao-Long Chen for useful discussions.
N.O. would like to thank the Particle Theory Group of the University
of Delaware for hospitality during his visit.
This work is supported in part by the DOE Grant
\#DE-FG02-91ER40626 (I.G. and Q.S.), GNSF grant 07\_462\_4-270 (I.G.),
and the Grant-in-Aid for Scientific Research from the Ministry of
Education, Science and Culture of Japan, \#18740170 (N.O.).


\end{document}